\documentclass[notoc]{JHEP3}
\usepackage{amssymb, epsfig}
\newcommand{\newc}{\newcommand}
\newc{\beqa}{\begin{eqnarray}}
\newc{\eeqa}{\end{eqnarray}}
\newc{\beq}{\begin{equation}}
\newc{\eeq}{\end{equation}}
\newc{\nonr}{\nonumber}
\newc{\hs}{\hskip 0.2cm}
\newc{\ra}{\rightarrow}
\newc{\ep}{\epsilon}
\newc{\tri}{\triangle}
\newc{\hl}{\hat{L}}
\newc{\hr}{\hat{R}}
\newc{\RE}{\mbox{Re}}
\newc{\IM}{\mbox{Im}}
\newc{\MeV}{\mbox{ MeV}}
\newc{\GeV}{\mbox{ GeV}}
\newc{\TeV}{\mbox{ TeV}}
\newc{\CKM}{\mathrm {CKM}}

\title{CP violation in 5D Split Fermions Scenario }
\author{We-Fu Chang\thanks{wfchang@triumf.ca}\ \  and John N. Ng \thanks{misery@triumf.ca}\\
TRIUMF Theory Group, 4004 Wesbrook Mall, Vancouver, B.C. V6T 2A3 CA }
\abstract{
We give a new configuration of split fermion positions in one extra
dimension with two different Yukawa coupling strengths for up-type, $h_u$, and
down-type, $h_d$, quarks at $\frac{h_u}{h_d}=36.0$. The new configurations can give enough CP violating (CPV) phase
for accommodating all currently  observed CPV processes. Therefore, a 5D standard model with split fermions
is viable.
In addition to the standard CKM phase, new CPV sources involving Kaluza-Klein(KK) gauge bosons
coupling which arise from the fact that unitary rotation which transforms
weak eigenstates into their mass eigenstates only holds for the zero modes which are the
SM fields and not for the KK excitations.
We have examined the physics of kaon, neutron, and $B/D$ mesons and found the
 most stringent bound on the size $R$ of the extra dimension
comes from $|\epsilon_K|$. Moreover, it depends sensitively on the width, $\sigma$, of the Gaussian wavefunction in the extra dimension used to describe
of the fermions.
When $\sigma/R \ll 1$ , the constraint will be lifted due
to GIM suppression on the flavor changing neutral current(FCNC)
and CPV couplings.
}
\preprint{CPX1028.tex\\ \today  \\ hep-ph/0210414}
\keywords{Extra Dimension, Fermion Masses,  CP Violation}

\begin{document}
\section{Introduction}
Recently, the prospect of large extra dimensions has been introduced\cite{ADD} as an alternative
view of  gauge hierarchy problem of the Standard Model(SM). The extra dimension scenario also
opens up new possible avenues for exploring physics beyond SM and many new degrees of freedom
can now be entertained.
 Following the same line a new geometrical
interpretation of the observed fermion mass hierarchy was introduced in  Ref \cite{AS}
by assuming that SM chiral fermions are localized at
different positions by a background potential in the extra dimensional space $y$. Depending
on the details of the model Gaussian or exponential wave functions in $y$
are for these zero modes.
The resulting effective 4-D Yukawa hierarchy can be viewed as the
overlapping wave functions between two different chiral fermions and is
exponentially suppressed as their relative distance $\tri y$. The further apart two chiral
fermions are the smaller their 4D Yukawa and hence a smaller mass for the fermion.

This setup will naturally generate
the tree level flavor changing coupling of Kaluza-Klein(KK) gauge
bosons, which can provide many interesting flavor changing neutral current effects. This has been
studied in detail for the charged leptons in \cite{TRIUMF}. The model used  is the 5D SM with
gauge and Higgs bosons all propagating in the bulk whereas the fermions are localized in the
fifth dimension with an unspecified potential. It is customary to
assume the fermions all have a  Gaussian
distribution in $y$, and one universal
Yukawa coupling to the bulk Higgs field. A realistic configuration was found
by \cite{MS} numerically that fits the observed quark masses and the quark mixing matrix
$V_{\CKM}$.
 Later on, it was pointed out by \cite{Branco} that
because the  structure zeros in the quark mass matrices of this particular solution the
resulting Jarlskog invariant \cite{Jarlskog} $J\equiv |Im(V_{ub} V_{cs} V^*_{us} V^*_{cb})|\sim 10^{-9}$
  is
four order of magnitude below the observed CP violation (CPV)
effects. The authors in \cite{Branco} went on and obtained  solutions that accommodate the observed CPV
phenomenon by extending to two  extra dimensions.

In this paper we point out that a minimum extension of the 5DSM  with different
Yukawa couplings $h_u$ for up-type and $h_d$ for down-type quarks has a
solution at $h_u/h_d = 36.0$ which reproduces observed mass spectrum
and the values of the CKM mixing angles in the left handed charged
current and at  the same time yields the desired strength of CPV effects.
Since the assumption of universal Yukawa coupling is made on account of simplicity
we deemed it worthwhile to explore the effects of relaxing this. Furthermore, it is not a priori
clear that a viable solution to quark masses and the CKM matrix can be found. Indeed
an extensive numerical scan is required just as in the case of \cite{AS}. This is described in
Sec 3.

The 5DSM is very rich in FCNC effects involving the KK excitations of the gauge and the Higgs
bosons \cite{DPQ} and \cite{TRIUMF}. New CPV couplings also accompany these excitations.
In addition to the one phase in $V_{\CKM}$ that operates on left-handed quarks there
are phases related to the rotation of the right-handed quarks. The latter rotation
is important since KK excitations of the gauge bosons except the W boson all couple
to the right-handed quarks. The same is true for the KK Higgs bosons.  Given that
there are severe experimental limits on these effects nontrivial constraints are expected.
It is our purpose to study in detail such constraints and how they impact the physics of
this class of models. After going  through all the consideration of current
experimental limits on 
FCNC and CPV, it is found that the kaon CPV parameter $|\epsilon_K|$
gives the most stringent limit on  $R$, the size of the
compactified extra dimension. However, the limit on $R$ strongly depends on the
ratio of $\rho \equiv \sigma/R$ where $\sigma$ is the Gaussian width of the fermions
which is basically a free parameter in this
model. For $\rho = 10^{-2}$ we obtain $R^{-1}>7\times 10^3\ TeV$ whereas for $
\rho=10^{-6}$ no meaningful bound on $R$ is found. This is due to the fact that as  $\rho $ becomes smaller
the Glashow-Iliopoulos-Maiani(GIM)\cite{GIM} mechanism is more operative and the FCNC and the
accompanying CPV effects are suppressed as in the SM.

This paper is organized as follows:
In sec.2, we first  outline the 5DSM model with split fermions. One bulk
Higgs field is employed to give masses to the zero mode fermions. It has
a constant profile in $y$. We are assuming that the localization of fermions is due either
to  addition bulk scalar fields  with nontrivial profiles or some other yet unknown mechanism.
The gauge boson are also bulk fields. Only
the necessary pieces of the effective 4D Lagrangian are given. For details
of gauge fixing and the derivation of the Feynman rules we refer to Ref.\cite{TRIUMF}.
For simplicity a  Gaussian profile with a universal width $\sigma$ is given to all the quarks.
This is a free parameter of the model and is natural for consistency of the model to have
$\sigma \ll R$. The fermion KK excitations in this scenario will be much heavier then those of
the gauge and Higgs bosons level by level and we shall not investigate them here.
In Sec.3 we describe the solution we found for the positions in $y$ of the chiral quarks. They
 are then used to calculate the rotation matrices  for the chiral quarks. This can be done up to
 some arbitrary phases.
In sec.4, we examine the FCNC and CPV constraint from kaon
physics. We also calculate the decay of $K^0 \rightarrow \pi \nu \bar{\nu}$ in
this model.  In sec.5, we show that the new physics discuss here will lead to violation of
weak universality.
The focus in Sec 6 is on how the new sources of universality violation and CPV
in this model on free neutron decays. This is important for next generation of cold neutron
experiments. The electric dipole moment of the neutron will also be discussed.
Sec 7 contains the conclusion. Finally the  needed Feynman rules are collected in
Appendix B


\section{5D SM Model with Split Quarks}
In \cite{TRIUMF} a  5DSM on the orbifold $S_1/Z_2$ with localized split leptons
and  Higgs and gauge bosons propagating in the fifth dimension
was constructed.
The derivation of  the effective 4D Lagrangian on the orbifold fixed point from the
5D bulk Lagrangian is given there  and is easily extended to the quark sector. We make an additional
assumption that the Yukawa couplings for the up type quarks ,$h_u$, and the down type quarks ,$h_d$,
are different.
After integrating out the fifth dimension and performing the
usual Kaluza-Klein decomposition of the 5D fields with the appropriate orbifold boundary
conditions, the 4-D effective  interaction of the  KK gauge boson with the zero mode  mass eigenstate
quarks can be summarized  as follows:
\beqa
{{\cal L}^{KK}\over \sqrt{2}}&=& -g_s G_\mu^{(n)}
\overline{q_i} \gamma^\mu T^a\left(U^{(n)}_L\hl + U^{(n)}_R \hr\right)
q_j\nonr\\
&& - e A_\mu^{(n)}
\overline{q_i} \gamma^\mu \left(U^{(n)}_L \hl + U^{(n)}_R \hr\right)
q_j\nonr\\
&& -  \left({g_2\over \cos\theta}\right)  Z_\mu^{(n)}
\overline{q_i} \gamma^\mu \left(g_L^q U^{(n)}_L \hl + g_R^q U^{(n)}_R \hr\right)
q_j\nonr\\
&& -{g_2\over \sqrt{2}}  W_\mu^{(n)}
\overline{u_i} \gamma^\mu U^{(n)}_L \hl d_j+ H.c.
\eeqa
where the $G_\mu^{(n)}, A_\mu^{(n)}, Z_\mu^{(n)}$ and
$W_\mu^{(n)}$ are the $n-$th KK excitations of the gluon, photon, $Z$
boson, and $W$ boson respectively. The chiral projection operators are $\hr=\frac{1+\gamma_5}{2}$ and
$\hl=\frac{1-\gamma_5}{2}$ and the family index is $i=1,2,3$. The rest are standard notations for the SM.
The matrices $U^{(n)}_{L/R}$ are a combination of the unitary transformations $V_{L/R}$ that takes
the weak quark eigenstates to their mass eigenstates
and the cosine weighting of the n-th KK modes. They are results after integrating
out  Gaussian distribution of localized
fermions in the fifth dimension. Explicitly, we have
\beq
\label{Umatrix}
U^{(n)}_L= V_L^\dag \left(\begin{array}{ccc}
  c_{n1}^L& 0&0 \\    0& c_{n2}^L &0  \\   0& 0 & c_{n3}^L\end{array} \right)
  V_L, \hskip 5mm
U^{(n)}_R= V_R^\dag \left(\begin{array}{ccc}
  c_{n1}^R&0 &0 \\   0 & c_{n2}^R &0  \\   0& 0 & c_{n3}^R\end{array} \right) V_R
\eeq
where the short hand notation $c_{ni}^{L/R}\equiv \cos (n y_i^{L/R} /R)$ and $y_i$ is the fixed
location in the fifth dimension of the quark $q_i$. Appendix B gives the detail
structure of the U matrices in component forms.

These $U$ matrices encode
FCNC information since they are not diagonal in the quark mass basis. Clearly when the fermions are not
split, i.e. all the $y_i=0$ these $U$ matrices become the identity matrix and the KK excitations
will not have FCNC couplings. This is a general feature of the split fermion scenarios that
gives rise to FCNC couplings for KK gauge boson to zero mode fermions. A second feature
of phenomenological importance is the presence of $U_R$ and now the KK gauge bosons are sensitive
to rotations of the right-handed quarks. Now we have effective extra $Z$ bosons
without explicitly adding another  gauge group.  Similarly there are a host of
extra Higgs bosons although we only have one Higgs doublet albeit it is a bulk filed.
Furthermore, the $U$ matrices also
contain additional CPV phases which are distinct from the $V_{\CKM}$ phase. We shall see later
that the positions $y_i$ can be expressed in units of $\sigma$ and some can be as far away as
few tens of $\sigma$ from the orbifold fixed point. It is reasonable to assume that $\rho$ to be
less than $10^{-2}$. From Eq. (\ref{Umatrix}) the strength of FCNC and CPV interactions are
controlled by $\rho$. For the first few KK states an expansion in $\rho$ is accurate and the
role of the GIM mechanism is evident.
We shall defer the discussion of this phenomenology  to later sections and instead turn our
attention to the quark mass matrices.

The quark  mass matrices stem from the interaction of fermions and the vacuum
expectation  value (VEV) of Higgs zero mode are given by
\beq
M^{U(0)}_{\{ij\}}=
-\frac{v_0 h_{u\{ij\}}}{\sqrt{2}}\exp\left[-\frac{\tri_{ij}^2}{4\sigma^2}\right],\hskip 4mm
M^{D(0)}_{\{ij\}}=-\frac{v_0
h_{d\{ij\}}}{\sqrt{2}}\exp\left[-\frac{\tri_{ij}^2}{4\sigma^2}\right],
\eeq
where $\tri_{ij}=|y_i-y_j|$, the distance between flavor $i$ and
$j$. They  are diagonalized by biunitary rotations  of matrices $V_{L/R}$
\beq
 M^U_{diag}=V_U^{L\dag} M^U V_U^R,\hskip5mm   M^D_{diag}=V_D^{L\dag} M^D V_D^R
\eeq
The  Yukawa couplings of the $n-$th KK Higgs, $n>0$, can be expressed explicitly as
\beq
h^{(n)}={g_2 \over M_W} V_L^\dag M^{(n)} V_R
\eeq
where the $M^{(n)}$ matrix is the convolutions of mass matrix and the weight
of $n$-th KK excitation
\beq
 M^{(n)}_{\{ij\}}= M_{\{ij\}}  \cos\frac{n \overline{y_{ij}^{LR}}}{R}
 \eeq
where $\overline{y_{ij}^{LR}}=(y^L_i + y^R_j)/2$. For completeness the 4D interaction
of the quarks with the KK Higgs boson is given by
\beq
-{\cal L}^{KK}_Y =h^{(n)}_{d\{ij\}} \overline{Q_i} H^{(n)} d^R_j +
             h^{(n)}_{u\{ij\}} \overline{Q_i} \tilde{H}^{(n)} u^R_j +
             H.c.
\eeq
where $H^{(n)}$ is the n-th KK Higgs doublet, $\tilde{H}^{(n)}=i\sigma_2
H^{(n)*}$, $Q_i$ is the SU(2) doublet quarks and  $u_i$ and $d_i$ are the up- and down-type SU(2) singlet
quarks respectively. The detail form of these effective Yukawa matrices is given in Appendix B. It suffices
to note that again FCNC and CPV interactions are present in the KK Higgs couplings and they
flip the chiralities of the quarks.

\section{A  New Solution }
In \cite{Branco}, the authors  analyzed the CP violating properties of
the solution obtained by \cite{MS}. They concluded that the resulting
CP violation is not enough to accommodate the observed experiments
in $K$ and $B$ systems so a total two extra dimensions are needed.
The solution given in \cite{MS} is obtained by assuming a universal Yukawa coupling
strength: $|h^{(0)}_{u\{ij\}}|=|h^{(0)}_{d\{ij\}}|=1.5$  which is chosen to be
bigger than unity so as to accommodate a separation
between the left-handed and right-handed quarks of the third family. It is also
claimed that no other solution was found. We have independently checked that this
indeed is the case.
Here we would like to investigate whether a solution can be found by giving the Yukawa
couplings a minimal flavor structure. Hence, we relax the assumption of universal Yukawa
coupling assumption. The minimum extension one can imagine is to allow two different
couplings for up- and down-type right handed quarks.
As in Ref. \cite{MS} a value of  $h_u=1.5 $ is used and  $h_d$ is now allowed to float.
The solution has to pass the requirement that the observed
mass spectrum and the CKM mixing,
which are summarized in Appendix A, can be reproduced.
Also the solution is required  to accommodate the experimental CPV
processes, namely the resulting value of the Jarlskog invariant is big enough.

The strategy for numerical searching is the  following: we start from
the solution given in \cite{MS}, which appears to be robust, and
then slightly vary $h_d$ from the
initial value, $1.5$, and let the program meander around the
initial configuration to find a set of new positions  for the fermions
which pass the two criteria mentioned above. Then the new position and
$h_d$ are used as initial condition for next iteration that $h_d$
is further driven away from $h_u$.
We found that the second criteria can be fulfilled only
when $h_u/h_d$ is larger then $33.0$ and we cannot find any
solution for $h_u/h_d>40.0$ which pass the first requirement.
As an example, we give one of the solutions at $h_u/h_d=36.0$,
namely  $|h^u_{\{ij\}}|= 1.5$
and $|h^d_{\{ij\}}|= 0.0417$ which satisfies all the
requirements and the averaged resulting Jarlskog invariant is $10^{-5}$.

\beq
Q_i=\sigma \left(\begin{array}{c}  0.0\\14.2349\\ 8.20333 \end{array}\right),
U_i=\sigma \left(\begin{array}{c}  6.13244\\ 20.092\\ 9.64483 \end{array}\right),
D_i=\sigma \left(\begin{array}{c} 19.4523\\ 5.15818\\ 10.1992\end{array}\right),
\eeq
In the notation of Ref. \cite{AS} the Gaussian width there is
$\mu^{-1}=\sqrt{2}\sigma$.
The solution was shifted in $y$-direction a little bit to make every fermion
position positive such that there is no conflict with the $S_1/Z_2$
compactification.

The corresponding mass matrices at the scale of $m_t$ are
\beqa
|M_U|&\simeq& \left(\begin{array}{ccc}
  .02056& 0 &  0 \\
  0 & 0.04694 & 1.28438\\
  85.226 & 0 & 148.112
\end{array}\right) \mbox{GeV},\\
|M_D|&\simeq& \left(\begin{array}{ccc}
  0 & 0.0087 &  0 \\
  0.0027 & 0 & 0.1178\\
  0 & 0.6909 & 2.553
\end{array}\right) \mbox{GeV}
\label{qmmatrix}
\eeqa
They give quarks masses (c.f. \cite{MS} )
\beq
\label{eq:masses}
 \begin{array}{cccc}
m_u(2\GeV)=& 2.40 \MeV,& m_c(m_c)= &1.39 \GeV\\
m_d(2\GeV)= &3.61 \MeV,& m_b(m_b)= &4.10 \GeV\\
m_s(2\GeV)=&60.20 \MeV,& m_t(m_t)= &170.9 \GeV\\
\end{array} \eeq

The resulting CKM matrix is
\beq
|V_{CKM}|= |(V^L_U)^\dag V^L_D|
\simeq \left(\begin{array}{ccc}
0.9748  &  0.2232 & 0.0018  \\
0.2230  &  0.9741 & 0.0364 \\
0.0099  &  0.0351 & 0.9993
\end{array}\right)
\eeq
 It is instructive  to compare Eq.(\ref{qmmatrix}) with the solutions of \cite{MS}.
Our $M_D$ has the same structure whereas $M_U$ is not diagonal.
The small elements in the $(23)$ and $(31)$ positions
 allow us to accommodate a CKM phase.
Unlike the SM, the coupling of gauge boson KK excitations in
general can be flavor and CP violating. There are $N_f(N_f\!+\!1)/2=6$, $N_f$ is the number
of family, phases for
each left-handed and right-handed rotation matrices $V_L$ and $V_R$.
One can use six quarks mass eigenstates to absorb $2N_f\!-\!1=5$ phases. So in total
there are seven physical complex phases remained. Therefore, there
should be seven corresponding Jarlskog-like invariant can be
constructed \cite{BR}. The six additional phases are associated with the rotation
of the right-handed zero modes which interact with the KK gauge and Higgs bosons.
Notice that the gauge group is not extended.
For the SM part, namely with zero mode fermions couple to the gauge boson zero modes,
the only CP violating source is in the one CKM phase.
One can check that the resulting SM Jarlskog
$J\equiv |Im(V_{ub} V_{cs} V^*_{us} V^*_{cb})|\sim 10^{-5}$
which is the right amount to produce the desired CP phenomenology.

Since we have no knowledge about where and how large  the additional
phases are, and  neither do we know the size of the parameter $\rho$,
a Monte Carlo survey of various combinations are performed. Then we use the
 very accurate experiments on kaon to give us the acceptable values.
This will be discussed in detail  the next section.
 However, we note here that certain combinations of
 the matrix elements of the $U$ matrices,
Table I, are very useful as they occur frequently in our calculations.
 We list them for  $ \rho=\{10^{-2}, 10^{-4}, 10^{-6}, 10^{-8} \}$.
The first four rows are related to $\tri S=2$ operators and the
others are related to $\tri S=1$ ones.
The numbers represent the maximum absolute values allowed.
For randomly chosen phases, the real and the imaginary parts
fluctuate  between  plus and  minus maximum absolute value.

\TABULAR[htb]{|c|c|c|c|c|c|}{
 \hline
 & $\rho=10^{-2}$ &  $\rho=10^{-4}$ &
  $\rho=10^{-6}$ &  $\rho=10^{-8}$ \\
 \hline
 $|\sum_{n=1}(U^{(n)L}_{sd})^2/n^2|$ & $2.4\times10^{-2}$& $3.3 \times 10^{-5}$ &
   $4\times10^{-13}$ & $4\times 10^{-21}$ \\
 $|\sum_{n=1} (U^{(n)R}_{sd})^2/n^2|$ & $1.9\times10^{-2}$ & $6.5\times10^{-5}$&
    $9.7\times 10^{-13}$ & $9.7\times 10^{-21}$ \\
 $|\sum_{n=1} (U^{(n)L}_{sd}U^{(n)R}_{sd})/n^2|$ & $1.5\times10^{-2}$ &$4.6\times10^{-5}$&
  $6.2\times10^{-13}$ &$6\times 10^{-21}$ \\
 $|\sum_{n=1} (y^n_{sd})^2/n^2|$ & $ 1.1\times 10^{-5}$ &$1.1\times10^{-8}$ &
  $1.3\times 10^{-16}$ &$1 \times 10^{-24}$\\
  \hline
$|\sum_{n=1} (U^{(n)L}_{sd}U^{(n)L}_{uu})/n^2|$ & $9.9 \times10^{-2}$ & $ 4.6\times 10^{-4}$&
   $4.9\times10^{-8}$ & $5\times10^{-12}$\\
$|\sum_{n=1} (U^{(n)L}_{sd}U^{(n)L}_{dd})/n^2|$ & $9.3\times10^{-2}$ & $ 4.5\times 10^{-4}$&
   $4.9\times10^{-8}$ & $5\times10^{-12}$ \\
$|\sum_{n=1} (U^{(n)L}_{sd}U^{(n)R}_{uu})/n^2|$ & $ 2.2\times10^{-3}$ & $ 2.2\times 10^{-4}$&
   $4.9\times10^{-8}$ & $ 5\times10^{-12}$ \\
$|\sum_{n=1} (U^{(n)L}_{sd}U^{(n)R}_{dd})/n^2|$ & $9.5\times10^{-4}$ &  $ 2.4\times 10^{-4}$&
  $4.9\times 10^{-8}$ & $5\times10^{-12}$ \\
$|\sum_{n=1} (U^{(n)R}_{sd}U^{(n)L}_{uu})/n^2|$ & $9.3\times10^{-2}$ &  $ 6.8\times 10^{-4}$&
   $7.6\times10^{-8}$ & $9\times10^{-12}$ \\
$|\sum_{n=1} (U^{(n)R}_{sd}U^{(n)L}_{dd})/n^2|$ & $8.8\times10^{-2}$ &  $ 6.7\times 10^{-4}$&
   $7.6\times10^{-8}$ & $9\times10^{-12}$\\
$|\sum_{n=1} (U^{(n)R}_{sd}U^{(n)R}_{uu})/n^2|$ & $2.3\times10^{-3}$ &   $ 3.4\times 10^{-4}$&
   $7.6\times10^{-8}$ & $9\times10^{-12}$ \\
$|\sum_{n=1} (U^{(n)R}_{sd}U^{(n)R}_{dd})/n^2|$ & $ 3.5\times10^{-3}$ &  $ 3.7\times 10^{-4}$&
  $7.6\times10^{-8}$ & $9\times10^{-12}$ \\
$|\sum_{n=1} (y^n_{sd}y^n_{uu})/n^2|$ & $3.9\times10^{-3}$ &  $ 2.5\times 10^{-5}$&
    $2.8 \times10^{-9}$ & $ 3\times 10^{-13}$ \\
$|\sum_{n=1} (y^n_{sd}y^n_{dd})/n^2|$ & $5.0\times 10^{-6}$ &  $ 1.0\times 10^{-7}$&
    $1.4\times 10^{-11}$ & $  1.4\times 10^{-15}$ \\
  \hline}
{Upper bounds of various summations }

One can learn from the result of our numerical experiment that
as the ratio $\rho$ goes down, the off-diagonal couplings
become smaller. This is to be expected because in the limit $\rho \ra  0$
the matrices $diag\{c^{L/R}_1, c^{L/R}_2, c^{L/R}_3\}\ra \mathbf{1}_{3\times 3}$
and the coupling matrices $U^{(n)}_{L/R}$ also reduce to a three by three identity matrix.
Note that in general the $\tri S=2$ transition is smaller
than $\tri S=1$ one due to the same reason.
Since $U$ controls the strengths of the off-diagonal transition they are  suppressed by GIM
mechanism as  $\rho \ll 1$. This relaxes the lower bounds on
$1/R$ that exists in the literature which  range from few TeV to few tens of
TeV  because $\rho$ is often taken to be a free but fixed parameter.
Physically, if one uses a background potential to localize the fermions  a small $\rho$ corresponds
to very a sharp kink.
How to arrange for such a potential and its stability is beyond the scope of this paper.
We note in passing that roughly speaking, the series sum can be expressed as $\rho^k$.
Depending on the combination of flavor and chirality, the index $k$ effectively varies from $1.5$ to $3.2$.

We give in Figure 1 the geography  of the quarks in the fifth dimension.
\FIGURE[t]{
\epsfxsize=320pt
\centerline{\epsfbox{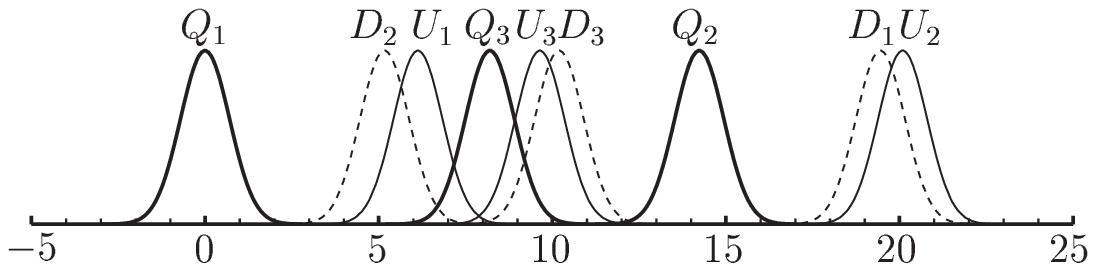}}
\caption{The map of fermions in the fifth dimension, in units of $\sigma$.}}

\section{Kaon phenomenology}
\subsection{$\mathbf{\tri M}$}

\FIGURE[ht]{
\epsfxsize=270pt
\centerline{\epsfbox{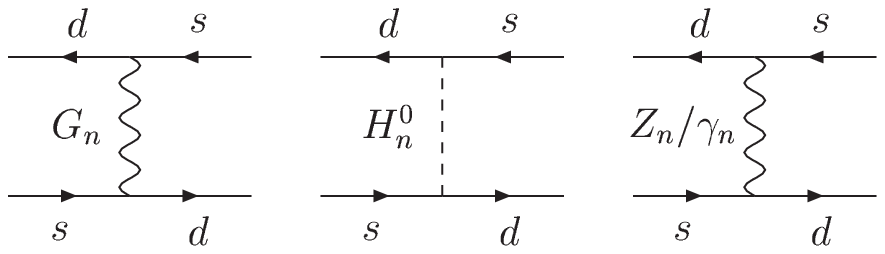}}
\caption{The dominant diagrams of KK exchange for  $\tri M$ }}

At the tree level, the flavor changing $\tri S=2$ transitions can
be mediated by exchanging neutral KK bosons, see Figure 2.
The effective $\tri S=2$ Lagrangian can be read
\beqa
{\cal L}_{\tri S=2}&=& \sum_n \frac{2g_s^2}{3n^2/R^2}\left[ U_{L\{ds\}}^n \bar{d}\gamma^{\mu}_L s+
U_{R\{ds\}}^n \bar{d}\gamma^{\mu}_R s\right]^2\nonr\\
&+&\sum_n \frac{1}{n^2/R^2}\left(\frac{e}{3}\right)^2\left[U^n_{L\{ds\}} \bar{d}\gamma^{\mu}_L s+
U^n_{R\{ds\}} \bar{d}\gamma^{\mu}_R s\right]^2\nonr\\
&+&\sum_n \frac{1}{n^2/R^2}\left(\frac{g_2}{\cos\theta}\right)^2\left[g_L U^n_{L\{ds\}} \bar{d}
\gamma^{\mu}_L s+
g_R U^n_{R\{ds\}} \bar{d}\gamma^{\mu}_R s\right]^2\nonr\\
&+&\sum_n \frac{1}{n^2/R^2}\left[h^n_{d\{ds\}} \bar{d_R}s_L+
h^{n*}_{d\{ds\}} \bar{d_L}s_R\right]^2.
\eeqa
Due to the larger coupling and color factors we can expect that the KK gluons dominate over
the others and give a contribution beyond the SM as given by
\[ \tri m_K\sim 2 \mbox{Re}M_{12} \sim 2\mbox{Re}M_{12}^{KK g},\hskip10mm
M_{12}=\langle K_0|H_{\tri S=2}|\bar{K^0}\rangle. \]
We using the standard method of vacuum insertion approximation(VIA) and the standard matrix
elements\cite{VIA},
\beqa
\langle K_0|(\bar{s}\gamma^\mu_L d )(\bar{s}\gamma_{\mu L} d )
|\bar{K^0}\rangle=\langle K_0|(\bar{s}\gamma^\mu_R d )(\bar{s}\gamma_{\mu R} d )
|\bar{K^0}\rangle= \frac13 m_K f_K^2,\\
\langle K_0|(\bar{s}\gamma^\mu_L d )(\bar{s}\gamma_{\mu R} d )
|\bar{K^0}\rangle= \left[\frac{1}{12}+\frac14\left({m_K \over m_s+m_d}\right)^2 \right]m_K
f_K^2.
\eeqa
Plug in the solution we found, Eq.(\ref{eq:masses}), and use  $m_K=497.6\MeV$\cite{PDG}
we have
\[
\langle K_0|(\bar{s}\gamma^\mu_L d )(\bar{s}\gamma_{\mu R} d )
|\bar{K^0}\rangle=45.86 \langle K_0|(\bar{s}\gamma^\mu_L d )(\bar{s}\gamma_{\mu L} d )
|\bar{K^0}\rangle.
\]
The chiral enhancement factor of $45.86$  differs from
\cite{tait} because our solution give a smaller  current  strange
quark mass.  However, it is easy to find an adjacent configuration which yields
a bigger strange quark mass  and given the uncertainty in the hadronic
calculation the difference is not serious. Thus, we find the contribution of
the KK gluons to the $K_L- K_S$ mass difference can be expressed as
\beq
\tri m_K= - \frac{2 g_s^2 R^2m_K f_K^2}{9}
\sum_{n=1}\frac{1}{n^2}\mbox{Re}\left[(U^n_{L\{ds\}})^2+(U^n_{R\{ds\}})^2
+91.72 U^n_{L\{ds\}} U^n_{R\{ds\}}\right],
\eeq
or the limit on $1/R$ as
\beq
1/R>1011\TeV \sqrt{\sum_{n=1}\frac{1}{n^2}\RE\left[(U^n_{L\{ds\}})^2+(U^n_{R\{ds\}})^2
+91.72 U^n_{L\{ds\}} U^n_{R\{ds\}}\right]},
\eeq
where the numerical inputs used are : $\alpha_s\sim 0.1$, $f_K=0.16$ GeV,
 $\tri m_K=3.488(79)\times 10^{-15}$ GeV
or $\tri m_K/m_K \sim 7\times 10^{-15}$  \cite{PDG}.

To a good approximation,
\beq
|\epsilon_K|\sim \frac{1}{2\sqrt{2}}\left|{\IM M_{12}\over \mbox{Re}M_{12}
}\right|.
\eeq
A similar expression from $|\epsilon_K|=2.282\times10^{-3}$ \cite{PDG}
gives constraint on $1/R$ as
\beq
1/R>12579\TeV \sqrt{\sum_{n=1}\frac{1}{n^2}\IM\left[(U^n_{L\{ds\}})^2+(U^n_{R\{ds\}})^2
+91.72 U^n_{L\{ds\}}U^n_{R\{ds\}}\right]}.
\eeq
We can put limits of $1/R$ by taking the maximum real or
imaginary part as inputs, as listed in Table 1 at various values of $\rho$  and
assume that the KK gluon accounts for most of the contribution to  $\epsilon_K$  and obtain
a stringent bound as seen in Table 2.  A caveat is in order.
We expect  the SM contribution for the CKM phase cannot
 be neglected. The relative size of the two contributions is not known. Hence, our
estimate is an optimistic one. Moreover, the hadronic uncertainties are large and our
bound is a good ball park number.

\TABLE{
\begin{tabular}{|c|c|c|c|}
 \hline
   \raisebox{-2ex}[0pt]{$\rho$} &
    \raisebox{-2ex}[0pt]{$\sqrt{L^2+R^2+2\times 45.86 L R}$}
     &\multicolumn{2}{c|}{$1/R$ (TeV) }\\
   \cline{3-4}
   && $\tri m_K$ & $|\epsilon_K|$ \\
  \hline
  $10^{-2}$ & $<1.17$ &  $>1200$ & $>14800$\\
   $10^{-3}$ & $<0.37$ &  $>370$ & $>4650$\\
   $10^{-4}$& $<6.6\times 10^{-2}$ & $>66$ & $>826$ \\
   $10^{-5}$& $<7.6\times 10^{-4}$ & $>0.8$ & $>10$ \\
  \hline
\end{tabular}
\caption{Lower bounds of $1/R$ from $\tri m_K$ and
$|\epsilon_K|$.}}

As  an order of magnitude estimation, we can easily extend of the
formulae for $\tri m_K$ to the cases of $D$ and $B$ mesons by
simply replacing $f_K\ra f_M$, $m_K\ra m_M$ and $m_K/( m_s+m_d)\ra m_M /
(m_Q+m_q)$, where  $Q$ and $q$ stand for the heavy and light
quark respectively which constitute the meson $M$.  Also the values of the $U$ matrix elements
are substituted by appropriate ones. Currently, these
quantities are less accurately known as their kaon counter part and
the constraint  they impose are much looser as seen in Table 3.

\TABLE{
\begin{tabular}{|c|c|c|}
\hline
  $\rho$ & $1/R(\tri m_D)$ & $1/R(\tri m_B)$ \\
  \hline
  $10^{-2}$ & $>34\TeV$ & $>3.9\TeV$ \\
  $10^{-3}$ & $>10.7\TeV$ & $>1.2\TeV$ \\
  $10^{-4}$ & $>2.0\TeV$ & $>0.3\TeV$ \\
  \hline
\end{tabular}
\caption{Bounds on $1/R$ from $\tri M_D$ and $\tri M_B$.}}

For the $\tri S=1$ transition, the important quantity to consider is $|\frac{\epsilon'}{\epsilon}|$.
In standard notation we have
\beqa
\left|{\epsilon'\over \epsilon}\right|= {\omega \over \sqrt{2}|\epsilon_K|}
\left({\mbox{Im}A_2 \over \mbox{Re} A_2}-{\mbox{Im}A_0 \over
\mbox{Re}A_0}\right),\\
A_I e^{i\delta_I}=<\pi\pi(I)|H_{\tri S=1}|K^0>,\nonr
\eeqa
where $\omega=\mbox{Re}A_2/\mbox{Re}A_0\sim 0.045$.
Again, we expect that the most important contribution comes  from KK gluon exchange.
Following the notation of Ref.\cite{buras}, we find that the dominant
contribution is from $Q_5=(\bar{d}\gamma^\mu_L s)\sum_q (\bar{q}\gamma_{\mu R}q)$
 and $Q_6=(\bar{d_i}\gamma^\mu_L s_j)\sum_q (\bar{q}_j\gamma_{\mu R}q_i)$,
 the QCD penguin in the SM, for the $\tri I=\frac12$
transition and $Q_7=\frac32(\bar{d}\gamma^\mu_L s)\sum_q e_q(\bar{q}\gamma_{\mu R}q)$,
$Q_8=\frac32(\bar{d_i}\gamma^\mu_L s_j)\sum_q e_q(\bar{q}_j\gamma_{\mu R}q_i)$,
 the electroweak penguin in the SM, for the $\tri I=\frac32$
 transition. Here the indices $i,j$ stand for color and
 also there are the operators $\tilde{Q}_i$ which are
 obtained from $Q_i$ by the exchange $L\Leftrightarrow R$.
Because parity is conserved in the strong interaction and assuming
the weak phase is negligible, we have
$\langle \pi\pi|Q_i|K^0\rangle=-\langle\pi\pi|\tilde{Q}_i|K^0\rangle$.
The dominant $\tri S=1$ effective lagrangian relevant to $Q_{5-8}$
due to KK gluon exchange is easy to compute and is given by
\beqa
&&{\cal L}^g_{\tri S=1}\nonr\\
&&= \sum_{{n=1 \atop q=u,d} } \frac{g_s^2 R^2}{n^2}\left[ U^n_{R\{ds\}} U^n_{L\{qq\}}(\bar{d}\gamma^R_\mu T^a s)
(\bar{q}\gamma^{\mu L} T^a q) + U^n_{L\{ds\}} U^n_{R\{qq\}}(\bar{d}\gamma^L_\mu T^a s)
(\bar{q}\gamma^{\mu R} T^a q) \right]\nonr\\
&&= \sum_{{n=1 \atop q=u,d}} \frac{g_s^2 R^2}{n^2}\left[ U^n_{R\{ds\}}
U^n_{L\{qq\}}
\left[\frac12(\bar{d}_i\gamma^R_\mu s_j)(\bar{q}_j\gamma^{\mu L} q_i)
-\frac16(\bar{d}\gamma^R_\mu s)(\bar{q}\gamma^{\mu L}q) \right]\right.\nonr\\
&&\left.\hskip25mm +U^n_{L\{ds\}}(U^n_{R\{qq\}}\left[\frac12(\bar{d}_i\gamma^L_\mu s_j)(\bar{q}_j\gamma^{\mu R}q_i)
 -\frac16(\bar{d}\gamma^L_\mu s)(\bar{q}\gamma^{\mu R} q)\right]
 \right].
\eeqa
Or we can rewrite it in the following form
\beqa
&&\sum_{n=1} \frac{g_s^2 R^2}{ 6 n^2}\left\{ U^n_{L\{ds\}}
\left[( U^n_{R\{uu\}}\!+\!2U^n_{R\{dd\}} )(Q_6\!-\!\frac13Q_5)
 + 2( U^n_{R\{uu\}}\!-\!U^n_{R\{dd\}})(Q_8\!-\!\frac13Q_7)\right]\right.\nonr\\
&&\left. + U^n_{R\{ds\}}
\left[( U^n_{L\{uu\}}\!+\!2U^n_{L\{dd\}})(\tilde{Q}_6\!-\!\frac13\tilde{Q}_5)
 +2(U^n_{L\{uu\}}\!-\!U^n_{L\{dd\}})(\tilde{Q}_8\!-\!\frac13\tilde{Q}_7)\right]\right\},
\eeqa
where the isospin symmetry breaking terms which are proportional to the
difference of the coupling matrices of up and down quarks can be seen explicitly.
Taking into account of the sign difference of $\langle Q_i\rangle$ and
 $\langle\tilde{Q}_i\rangle$,
we computed the  dominant contribution to be given by
\beqa
&&\left|{\epsilon'\over \epsilon}\right|
\simeq {\omega \over \sqrt{2}|\epsilon_K|\RE A_0}\nonr\\
&&\times\left\{ \left.\sum_{n=1}{g_s^2 R^2 \over 6 n^2}\IM\right[
\left[ U^n_{L\{ds\}}( U^n_{R\{uu\}}+2U^n_{R\{dd\}} )
 -U^n_{R\{ds\}}( U^n_{L\{uu\}}+2U^n_{L\{dd\}} ) \right]\right.\nonr\\
&& \times\left(\langle Q_6\rangle_0-\frac13\langle Q_5\rangle_0\right)
-\frac{2}{\omega}\left[ U^n_{L\{ds\}}( U^n_{R\{uu\}}-U^n_{R\{dd\}} )
 -U^n_{R\{ds\}}( U^n_{L\{uu\}}-U^n_{L\{dd\}} ) \right]\nonr\\
&&\left.\left.  \times\left(\langle Q_8\rangle_2-\frac13\langle
Q_7\rangle_2\right)\right]\right\}.
\eeqa

If we use  the observed value of $|\epsilon'/\epsilon|=(1.8\pm0.4)\times
10^{-3}$\cite{PDG}, the bound for $R$ is very weak. Instead with the aid of
 VIA and the limits of $1/R$ derived from $|\epsilon_K|$,
an upper bound on  $|\epsilon'/\epsilon|$ from KK gluons can be predicted to be
$\{3.4\times 10^{-8},3.4\times 10^{-8},2.3\times 10^{-8},1.1\times 10^{-6} \}$
at $\rho=\{10^{-2},10^{-3},10^{-4},10^{-5}\}$.
Hence, for this model the direct CPV in kaon decays comes from the CKM phase and
{\em not} the KK gluons or other gauge and Higgs excitations.

\subsection{$\mathbf{K^+\ra \pi^+ \nu\bar{\nu}}$ and $ \mathbf{K_L\ra \pi^0 \nu \bar{\nu}}$}
The two neutrinos semileptonic rare kaon decays are very sensitive probes of
physics beyond the SM since they are relatively clean theoretically. Their discussion will involve
the lepton configuration in $y$. Various kinds of constraint
on the lepton positions in the extra dimension have been
discuss \cite{TRIUMF}. Here our main purpose is to examine the contributions from the
quark configurations we found . So we will take a simple set in which
the lepton mass matrix is diagonal and use it as an example to evaluate the limit we can put on
the size $R$ of the new physics. This is sufficient for now. A complete job will
require the knowledge of the values of the neutrino masses and the leptonic CKM matrix which
are lacking. We are justified to  neglect the question of neutrino mass
in the extra dimension scenario. The interested reader can consult \cite{bulknu}
and references therein. Without further ado
we adopt the lepton positions in the fifth dimension given by \cite{MS} and
shift them the same amount as we did for quarks:
\beq
L_i=\sigma \left(\begin{array}{c}  22.9943\\ 8.7461\\ 7.3319 \end{array}\right),
E_i=\sigma \left(\begin{array}{c}  15.7429\\ 14.3287\\2.8774
\end{array}\right).
\eeq
To a good approximation, the left-hand and right-hand rotation
matrices for lepton can be treated as identity matrices. The main contribution now arise from KK $Z$
boson exchange, see Figure 3. Their couplings will simply be the SM coupling multiplied by
the cosine weighting $\sqrt{2} c_n^{L/R}$ for the $n$-th KK boson.
\FIGURE[ht]{
\epsfxsize=100pt
\centerline{\epsfbox{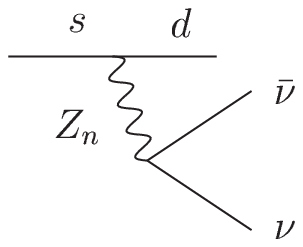}}
\caption{The dominant KK diagram
for  $K\ra\pi\nu\nu$ decays }}

We now have all the ingredients to compute the effective Hamiltonian for the rare decay. It is
given by
\beqa
H_{eff}&=&{2g_2^2 g_L^s g_L^\nu\over \cos^2\theta_W}
\sum_{{n=1\atop l=e,\mu,\tau}} {U^n_{L\{sd\}} U^n_{L\{\nu_l\nu_l\}}\over n^2/R^2 }
(\bar{s}\gamma^\mu_L d)(\bar{\nu_l}\gamma_{L\mu} \nu_l)\nonr\\
&+&{2g_2^2 g_R^s g_L^\nu\over \cos^2\theta_W}
\sum_{{n=1\atop l=e,\mu,\tau}} {U^n_{R\{sd\}} U^n_{L\{\nu_l\nu_l\}}\over n^2/R^2 }
(\bar{s}\gamma^\mu_R d)(\bar{\nu_l}\gamma_{L\mu} \nu_l)
\eeqa
and we have adopted the notations of Buras\cite{buras} and extended
the formulae to include new right-handed operators
The branching  ratio can be written in the following form
\beqa
Br(K^+\ra \pi^+\nu\bar{\nu})\simeq k_+\left(\left|{X_L\over\lambda^5 }\right|^2
+\left|{X_R\over\lambda^5 }\right|^2\right),\\
Br(K_L\ra \pi^0\nu\bar{\nu})\simeq k_L\left[\left({\IM X_L \over \lambda^5}\right)^2+
\left({\IM X_R \over \lambda^5}\right)^2\right],
\eeqa
where
\beq
X_{L/R}= \frac{8\pi\sin^2\theta_W}{\alpha\cos^2\theta_W}
{ g_L^\nu g_{L/R}^s  R^2M_W^2 \over V_{ts}^* V_{td}}
\sum_{n=1} {U^n_{L/R\{sd\}} U^n_{L\{\nu\nu\}}\over n^2 }.
\eeq
The various quantities are  $k_+=4.57\times 10^{-11}$,
 $k_L=k_+ \tau(K_L)/\tau(K^+)=1.91\times10^{-10}$, and $\lambda= |V_{us}|$.
 Not surprisingly the FCNC
couplings are analogous to those in  the $|\ep_K|$ calculation.
We can  plug in the constraint on $1/R$ from before  and get the
following prediction:
$Br(K^+\ra \pi^+\nu\bar{\nu})<\{ 2.7\times10^{-17}, 3.3\times10^{-17},
3.6\times10^{-17}, 3.7\times10^{-12}\}$
 at $\rho=\{10^{-2}, 10^{-3}, 10^{-4}, 10^{-5}\}$.
This is to be compared to the SM prediction of $Br(K^+\ra \pi^+ \nu \bar{\nu})|_{SM}=(.75\pm .29) \times
10^{-10}$ \cite{Buras2}.
The result is very interesting compared to the present experiment
value $Br(K^+\ra \pi^+\nu\bar{\nu})=(1.5{+3.4\atop
-1.2})\times10^{-10}$\cite{adler}.
$X_L$ and $X_R$ are essentially the amplitudes of the new physics. $X_L$ is expected to interfere
with the SM amplitude. Since we do not know the sign of the new phases relative to the CKM phase
the numbers given only indicative and not robust numbers.
As one can see, when $\rho=10^{-5}$ the KK contribution can be as large as twenty percent
of SM value at the amplitude level which will modify the branching
ratio up to fifty percent. Even at $\rho=0.01$, the branching
ratio modification could also reach  $0.1$ percent level.
It will be interesting to have more precise experimental bound.
If the experimental result persists to be higher
than the SM prediction it could signal a positive contribution from the mechanism we
propose here.

A crude  upper limit for $K_L\ra\pi^0\nu\bar{\nu}$ decay can be
obtained by assuming the maximum allowed phases and simply multiplying a factor $k_L/k_+=4.17$
to the above prediction.

It will be difficult to obtain  bounds or  upper limits of the lepton flavor violation in
$K$ decays, for instance the processes $K^+\ra\pi^+e^- \mu^+$, $K^+ e^+\mu^-$,
$K_L\ra\mu e$ and $K_L\ra \pi^0 e \mu$. Besides the strong
dependence  on the exact configuration of the lepton sector  they  also crucially depend
on the leptonic CKM matrix which currently have no information. Hence, they are
best left for future studies.

\section{Universality of $\mathbf{\pi\ra e\nu, \mu\nu}$}
\FIGURE[ht]{
\epsfxsize=200pt
\centerline{\epsfbox{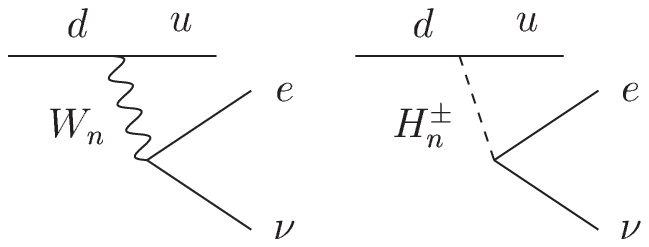}}
\caption{Dominant KK contributions
for  neutron beta decay and $\pi^\pm$ decays }
}
The universality test of pion leptonic decays is a cornerstone
for the SM. In this model, the interaction of the physical $W$
is same as in the SM. So the universality tests using leptonic channels  of physical $W$
decay will not be altered from the SM prediction in this model \cite{TRIUMF}.
However, the $\pi$ decay
will be modified by exchanging the virtual  KK excitations as depicted in Figure 4.
Since the KK Higgs couplings to the fermions are suppressed by the
light lepton masses we expect that the KK $W$ exchange gives  the dominant contribution.
The resulting modification of decay rate ratio must satisfy the current
universality test bound \cite{PDG}, i.e.
 \beq
\tri\left({\Gamma(\pi\ra e\nu)\over\Gamma(\pi\ra \mu\nu) }\right)_{KK}
\sim M_W^2R^2\sum_{n=1}\frac{1}{n^2}\mbox{Re}
 \left[ U^{L(n)}_{ud}\left( U^{L(n)*}_{ee}-U^{L(n)*}_{\mu\mu}\right )\right] \lesssim
 10^{-7}.
 \eeq
This is easily satisfied by the
constraint  on $R$ and $\rho$ derived from $|\epsilon_K|$. The universality violation will only
be changed by the amount of $\{10^{-11}, 10^{-11}, 3\times10^{-11}, 2\times10^{-9}\}$
for  $\rho=\{10^{-2},10^{-3},10^{-4},10^{-5}\}$.

\section{Neutron $\mathbf{\beta}$ decay }
Following the notation of \cite{Jackson}, the most general
differential  rate of a free neutron decays into proton plus electron and
neutrino can be expressed as
\beqa
d^2\Gamma&\propto& E_e|\overrightarrow{p_e}|(E_e^{max}-E_e)^2
dE_e d\Omega_e
d\Omega_\nu\nonr\\
&&\times\left[1+ a{\overrightarrow{p_e}\cdot \overrightarrow{p_\nu} \over E_e E_\nu}
 +b {m_e \over E_e}
+ A {P\cdot\overrightarrow{ p_e} \over E_e}
+ B {P \cdot \overrightarrow{p_\nu}\over E_\nu}
+D{P\cdot(\overrightarrow{p_e}\times \overrightarrow{p_\nu})\over E_e
E_\nu}\right],
\eeqa
where $P$ is the polarization vector of neutron.
In the SM, neutron beta decay is solely mediated by
virtual $W$ boson exchanging.
Neglecting the recoil correction, the coefficients
are also given by \cite{Jackson}
\beq
a={1-\lambda^2\over 1+3\lambda^2},\hskip2mm
b=0,\hskip2mm
A= 2{\lambda(1-\lambda)\over 1+3\lambda^2},\hskip2mm
B=2{\lambda(1+\lambda)\over 1+3\lambda^2},
\eeq
with $\lambda\equiv |g_A|/|g_V|$
and the CP violating triple correlation coefficient $D$
is small.
In principle, $\lambda$ can be calculated by lattice QCD
from the first principle and it has been precisely measured
as $1.267(3)$ \cite{PDG} which agree well with the measurements
on the coefficients $a, A, B$.

In addition to the exchange of SM $W$ boson, in the split fermion scenarios the neutron
beta decay also receive contribution of KK $W$ and KK charged Higgs
boson, see Figure 4. The new  four Fermi interaction term is
\beqa
H^\beta_{int}&=& \sum_{n=1}
\frac{g_2^2}{2n^2/R^2} U^n_{L\{du\}} U^n_{L\{e\nu\}}(\bar{u}\gamma_{L\mu}
d)(\bar{e}\gamma_L^\mu \nu)\nonr\\
&+&\sum_{n=1}
\frac{1}{n^2/R^2} h^{n+}_{\{e\nu\}}
\bar{u}[ h^{n+}_{S\{ud\}} -\gamma^5 h^{n+}_{P\{ud\}}]d[\bar{e_R}\nu_L] + H.c.\\
&\equiv& {g_2^2 V_{ud}\over 2M_W^2}\left[a_W(\bar{u}\gamma_{L\mu}
d)(\bar{e}\gamma_L^\mu \nu) + a_S(\bar{u} d)(\bar{e}\nu)+
a_P(\bar{u}\gamma^5 d)(\bar{e}\nu) \right] + H.c.,
\eeqa
where $h^{n+}$s are the coupling of fermions and KK charged Higgs
bosons (refer to Appendix B for their couplings).
In general, $a_W$, $a_S$ and $a_P$ are complex.
Now it is interesting to examine  how the new interaction will affect
the neutron decay, especially there are new CPV source in the
KK $W$ couplings and the presence of additional scalar interaction.

First, we note that the present of $a_W$ and $a_H$ will alter the
total decay rate of neutron or the definition of $G_\beta$
which will be discussed in the next subsection on $\tri r_\beta-\tri r_\mu$.
We now look at effect of $a_W$ more closely.
At first glance it appears that there is a contribution to  the CPV coefficient $D$.
But from \cite{Jackson}
\beqa
D\simeq  \frac\xi2 \IM\left({\IM g_V \over g_V}-{\IM g_A \over
g_A}\right),\nonr\\
\xi=4\delta_{J_n,J_p}\left({J_n\over J_n+1}\right)^{1/2}{\lambda M_F M_{GT}
\over (1+|a_S|^2)|M_F|^2+ \lambda^2|M_{GT}|^2},
\eeqa
where $M_F$ and $M_{GT}$ are the Fermi and Gamow-Teller nuclear
matrix elements, for neutron beta decay $|M_F|=1, |M_{GT}|=\sqrt{3}$. And $J_n, J_p$ are the spin of neutron and
proton respectively. Since KK charged current gives the  same contribution to
$g_V=g_V^{SM}(1+a_W)$ and $g_A=g_A^{SM}(1+a_W)$ at the quark level, unless the CPV be generated
through long distance physics, we conclude there is no new effect on the
CPV coefficient $D$ in neutron beta decay.
Also because the  absence of tensor coupling at  the tree level, the
coefficients $A$ and $B$ will not be altered.
The new scalar interaction will modify the $a$ and $b$
parameters and  are given by
\beqa
a={(1-|a_S|^2)|M_F|^2 -\lambda^2 |M_{GT}|^2/3
\over
(1+|a_S|^2)|M_F|^2 +\lambda^2 |M_{GT}|^2 } ,\nonr\\
b={2|M_F|^2 \RE( a_S)
\over
(1+|a_S|^2)|M_F|^2 +\lambda^2 |M_{GT}|^2 } ,
\eeqa
with the current experimental data $a=-0.102(5)$ \cite{PDG}, $a_S$
is bounded by $|a_S|^2<0.138$ and related to $b$ as
\beq
a_S={\sqrt{2}R^2 \over 4 G_F V_{ud}}\sum_{n=1}{h^{n+}_{S\{ud\}} h^{n+}_{\{e\nu\}} \over
n^2},\hskip2mm
b={2\RE( a_S )\over 5.82+|a_S|^2} .
\eeq
One immediately see that independent of $\rho$, $b$ has to be in the
range of  $\pm 0.124$.
The prediction of $a_S$ is very small, $<10^{-14}$,
due to the suppressed of electron mass. There is no hope to
constraint this model by precision test of neutron beta decay.

\subsection{$\mathbf{\tri r_\beta -\tri r_\mu}$}
\FIGURE[ht]{
\epsfxsize=300pt
\centerline{\epsfbox{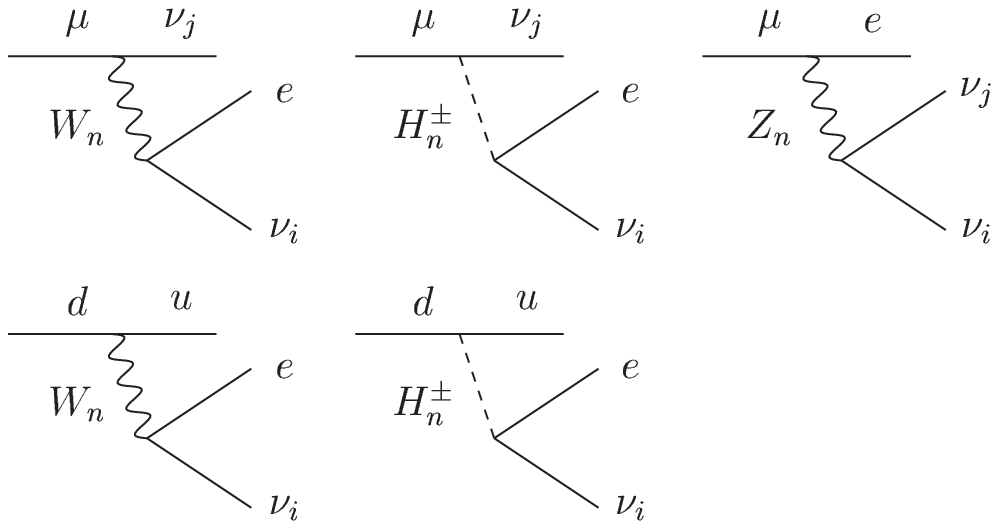}}
\caption{Tree level KK
contributions to $\tri r_\beta- \tri r_\mu$}}

It is well known that many new physics will modify the effective four Fermi coupling \cite{TRIUMF}
and \cite{Mar}. This holds
true for the split fermions scenario in general, see Figure 5 for the Feynman diagrams.
We define the process dependent Fermi constant $G_{\mu}$ as
\beqa
\label{gmu}
{G_{\mu}\over \sqrt{2} }&\equiv& {g^2_2 \over 8M_W^2 }\left[1+\tri
r_\mu\right]\nonr\\
&=&{g^2_2 \over 8M_W^2 }
\left[\left(1+R^2M_W^2(a_1+a_3)\right)^2
+\left({a_2+2a_4\over 2}\right)^2R^4M_W^4\right]^{\frac12}.
\eeqa
where the coefficients $a_{i,j,k}$ are the result of summing  over all neutrino species and
explicitly given by
\beqa
a_1\sim \sum_{i,j,n=1} { U^n_{L\{j\mu\}}U^{n*}_{L\{ei\}} \over
n^2},\hskip2mm
a_2\sim \sum_{i,j,n=1} { 4 h^n_{\{j\mu\}}h^{n*}_{\{ie\}} \over g^2_2 n^2},\nonr\\
a_3\sim \left({2 g_L^\nu g_L^e \over \cos^2\theta }\right)
\sum_{i,j,n=1} { U^{n}_{L\{e\mu\}}U^{n*}_{L\{ij\}} \over n^2},\hskip2mm
a_4\sim\left({2 g_L^\nu g_R^e \over \cos^2\theta }\right)
\sum_{i,j,n=1} { U^{n}_{R\{e\mu\}}U^{n*}_{L\{ij\}} \over n^2 }.
\eeqa
The square bracket in Eq.(\ref{gmu}) gives the modification to the SM Fermi coupling
constant, $G_{SM,F}=\sqrt{2}g^2_2/8M_W^2$ and also generalizes the usual KK result\cite{Mar}.
We have a similar result for neutron beta decay:
\beq
\label{gn}
{G_{\beta}\over \sqrt{2} }
={g^2_2 V_{ud} \over 8M_W^2 }
\left[\left(1+R^2M_W^2 a^n_1\right)^2
+\left({a^n_2\over 2}\right)^2R^4M_W^4\right]^{\frac12},
\eeq
where
\[
a_1^n \sim \sum_{i,n=1} { U^{n}_{L\{ud\}}U^{n*}_{L\{ei\}} \over
V_{ud}n^2},\hskip2mm
a_2^n \sim \sum_{i,n=1} { 4 h^n_{\{ud\}}h^{n*}_{\{ie\}} \over V_{ud} g^2_2 n^2}.
\]
They are due to the KK $W$ and $H^\pm$ exchange.
Since the new physics modifies the effective Fermi couplings differently and this is
expressed in the following
form
\beq
G_F^\beta =G_\mu V_{ud} (1-\tri r_\mu +\tri r_\beta).
\eeq
Taking the previous diagonal lepton configuration, namely no FCNC
in lepton sector, and expand the above formulas up to ${\cal O}(M_W^2R^2)$,
\beq
\left(\tri r_\mu- \tri r_\beta\right)_{KK}\sim
2 R^2M_W^2 (a_1^\mu- a_1^n).
\eeq
With the inputs from kaon decays we  predict the upper limits of $\left(\tri r_\mu-\tri r_\beta\right)_{KK}<\{
6\times10^{-12},8\times10^{-11},2\times10^{-9},2\times10^{-5} \}$
for $\rho=\{10^{-2},10^{-3},10^{-4},10^{-5}\}$. Hence, violation of universality in these
channels are expected to be small.

\subsection{Neutron EDM}
\FIGURE[ht]{
\epsfxsize=200pt
\centerline{\epsfbox{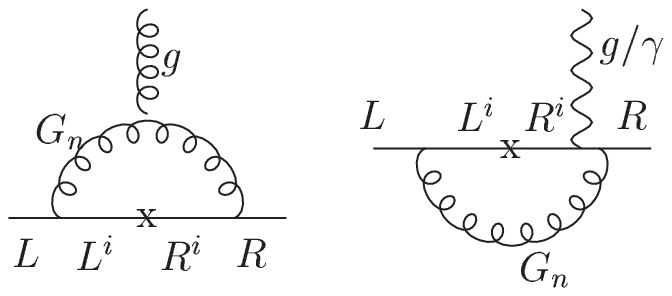}}
\caption{One loop diagrams for the chromo-dipole
moment of a quark.}}
The neutron EDM is one of the most stringent test of CPV in the flavor conserving sector.
Using the SU(6) quark model, the neutron EDM can be related to related to u(d) quarks'
EDM $d_u(d_d)$ and their chromoelectric dipole moments  $d_u^g(d_d^g)$ as follows
\beq
\left(\frac{D_n}{e}\right)= \frac43\left(\frac{d_d^\gamma}{e}\right)
-\frac13 \left(\frac{d_u^\gamma}{e}\right)
+\frac49\left(\frac{d_d^g}{g_s}\right)+\frac19 \left(\frac{d_u^g}{g_s}\right).
\eeq
As an estimate of the size of the new effects we assume that the chromoelectric dipole
moment to  be the dominating contribution to neutron EDM, see Figure 6. We can ignore
the Weinberg three gluon operator for now.
Without going into the details of  the calculation  the one-loop transition amplitude
can be estimated to be:
\beqa
\label{eq:nedm}
\left( {d^g_u \over g_s}\right)\sim \sum_{{ n=1\atop q= c, t}}{g_s^2 m_q \over 4\pi^2 M_n^2}
 \IM( U^n_{R\{qu\}} U^n_{L\{qu\}} ),  \nonr\\
\left( {d^g_d \over g_s}\right)\sim \sum_{{ n=1\atop q=s,b}}{g_s^2  m_q \over 4\pi^2 M_n^2}
 \IM( U^n_{R\{dq\}} U^n_{L\{dq\}}).
\eeqa
To make a dipole operator, one need at least one mass insertion either on the
internal or external fermion line to flip the chirality.
Note that the light quark, $u, d$, contribution are absent
because the flavor diagonal coupling is real.
This  is easy to see
by looking at the component form  of the coupling matrix, the
complex numbers come in conjugate pair and the KK weighting factors $c_n^{L/R}$
are also real.

The contribution from the third family is enhanced by their masses
but suppressed by the smallness of off-diagonal coupling between
the first and third family.
The factor $1/4\pi^2$ is  due to the loop factor and
$M_n^2=n^2/R^2$ is from the propagator of KK  gluon.
By using the Eq.(\ref{eq:nedm}) and the lower bound obtained from
$|\epsilon_K|$, the predicted limits of neutron EDM are
 $d_n<\{6.6\times10^{-33}, 1.6\times10^{-33},2.6\times10^{-35}\}$(e-cm)
for $\rho=\{10^{-2}, 10^{-3}, 10^{-4}\}$. Compared with the
current experimental bound, $|d_n|<6.3\times10^{-26}$e-cm \cite{nEDM}.
This is even smaller then the expected SM contribution.

\section{Conclusion}

\TABULAR[ht]{|c|c|c|}{
\hline
 Observable & Constraint & Dominate KK\\
 \hline
$\tri m_K$ & $\RE[U^2_{L\{sd\}} +U^2_{R\{sd\}}+91.72U_{L\{sd\}}U_{R\{sd\}} ]$ &
$G_n$\\
$|\epsilon_K|$& $\IM[U^2_{L\{sd\}} +U^2_{R\{sd\}}+91.72U_{L\{sd\}}U_{R\{sd\}} ]$ & $G_n$\\
$|\ep'/\ep|$& $\IM[U_{L/R\{sd\}}U_{R/L\{qq\}}]$ & $G_n$\\
$K^+\ra\pi^+\bar{\nu}\nu$ & $|U_{L\{sd\}}U_{L\{\nu\nu\}}|^2+|U_{R\{sd\}}U_{L\{\nu\nu\}}|^2$ & $Z_n$\\
$K_L\ra\pi^0\bar{\nu}\nu$ & $\IM[U_{L\{sd\}}U_{L\{\nu\nu\}}]^2+\IM[U_{R\{sd\}}U_{L\{\nu\nu\}}]^2$ & $Z_n$\\
\hline
$\tri m_D$ & $\RE[U^2_{L\{cu\}} +U^2_{R\{cu\}}+5.5U_{L\{cu\}}U_{R\{cu\}} ]$ &
$G_n$\\
$\tri m_B$ & $\RE[U^2_{L\{bd\}} +U^2_{R\{bd\}}+5.2U_{L\{bd\}}U_{R\{bd\}} ]$ &
$G_n$\\
\hline
${ Br(\pi\ra\nu e)\over Br(\pi\ra\mu\nu)}$& $\RE[U_{L\{ud\}}(U_{L\{ee\}}-U_{L\{\mu\mu\}})]$
& $W^\pm_n$\\
\hline
$n\ra p e\bar{\nu}$ & $ h^\pm_{S\{ud\}}h^\pm_{e\nu}$ &$H^\pm_n$\\
$\tri r_\beta-\tri r_\mu$ & $|U_{L\{ud\}}U_{L\{e\nu\}}|^2-
 |U_{L\{\mu\nu\}}U_{L\{e\nu\}}|^2 $ & $W^\pm_n$\\
$d_n$ & $\IM [\sum_{q=c,t} m_q U_{R\{qu\}}U_{L\{qu\}} ]
+4\IM[ \sum_{q=s,b} m_q U_{R\{qd\}}U_{L\{qd\}} ]$ & $G_n$ \\
\hline
}{Summary of constraint from phenomenology. It is understood the given
expression actually is the sum over all contribution from KK excitation.
Refer to the text for the accurate formulae. }

We have succeeded in finding  a realistic 5D split fermion
model,that yields the observed masses spectrum and
give correct amount of CKM CPV phase. This is important since KK excitations alone
will not account for $\epsilon'/\epsilon$.
We achieved this by bestowing  different Yukawa couplings for the up
quarks, $h_u$, and for the down-quarks, $h_d$. We have  fixed $h_u$ at
the value of $1.5$ and vary $h_d$.
Our numerical search indicates that the
above requirements can be fulfilled only in a nontrivial and narrow window
of $h_d$, $33.0<h_u/h_d<40.0$. What we presented is an existence proof but not
necessary a unique solution. It also shows  a moderate amount of tuning
in the flavor structure but not excessive.

There are interesting new  CPV and FCNC phenomenology in this model, i.e. the
CPV and FCNC coupling of KK-excitations and SM fermions
can be generated naturally  at the tree level due to the fact that
in the effective 4D theory the Yukawa matrix of KK modes receive
different weighting from the SM mass matrix.
The CPV characteristic of this model is unique in that there are CPV
interactions in the neutral current sector that are of the $(V+A)$ and $(V-A)$ type. They arise from
 KK photons, KK $Z$s, and KK gluons. Additional  CPV  coupling
for $(V-A)$ charged current from KK $W$ and  $S$-type and $P$-type couplings from
 the KK neutral and charged Higgs bosons are present. There are No $(V+A)$ vector
charged current in the model.
Unlike the SM CKM phase, it has six extra complex phases
arising from rotations of the $SU(2)$ singlet right-handed quarks.
In principle they are measurable once KK excitations are found.
Obviously, to determine all of them we need at least seven
linearly independent experiments to ping down all the phases.

To date, CPV effects are only observed in the Kaon and B system. In this scenario both
the CKM phase and new physics can come into play. The relative amount of the
two contributions are not known. We made the assumption that the new physics is at
least as large as the SM in $\tri m_K$ and $|\ep_K|$. If the KK gluon contributions
are only a fraction of the SM then the constraint on $R$ is even stronger provided
no fortuitous cancellations among elements of $U$.

From our discussions  one can see that the most stringent constraint on
$R$ the extra dimension size comes from the kaon CPV parameter
$|\ep_K|$. Depending  on $\rho$ the conservative lower bound on $1/R$
ranges from $10^{4}$ TeV at $\rho=10^{-2}$ to $10$ TeV at
$\rho=10^{-5}$. If by accident the imaginary parts of the contribution from
summing over KK excitation is nearly vanish, the real part, from $\tri m_K$,
will still give a strong constraint although one order of magnitude weaker. In Table 4
we summarize the constraints from other decays we have examined.
The rare decays of the kaon is next most sensitive and they probe a different source,
 i.e. the KK $Z$ and thus are very complementary.

The implications of our results to collider physics depends crucially on $\rho$.
The value of $\rho =10^{-2}$ is not unnatural but lead to KK excitations in the mass range
of $10^3$ to $10^4$ TeV which is out of reach for the foreseeable future. On the other hand
if $\rho<10^{-6}$ then TeV KK gauge bosons
production can be expected. Moreover, the KK fermion will be completely out of reach
since we expect the masses to be govern by $1/\sigma$. This is contrast to models
which has all SM particles in the bulk\cite{UED}.

To sum up, we have shown that it is still viable to have a 5D split fermion model
to explain both the flavor and hierarchy problem. However,
 extra dimension with size larger then $(\mathrm{TeV})^{-1}$ will require a dynamical model which
confines the fermions in a very small region in the extra dimension. The rich
CPV and FCNC phenomenology of the model makes it interesting to pursue new rare kaon
decay experiments as well as B and D decays since they probe different sectors of the
model.

Although we have focused on one specific model with Gaussian wavefunctions the feature
of the tree-level FCNC and CPV couplings in KK sector is generic for this class of models
with nontrivial fermion profiles. This is true also
 for models built with multi-located  fermions(or branes)\cite{multi} regardless
of the number of extra dimensions and the exact shape of how they
spread in the extra dimension(s). Our analysis can be carried out in these models
and similar tight constraints on these models are expected.

\acknowledgments
 This work is supported in part by the Natural Science and
Engineering Council of Canada.
WFC wish to thank the Institute for Nuclear Theory at the University of
Washington for its hospitality during the completion of this work.
\newpage
\appendix
\section{SM parameters}
In the search of  new configuration, we used the
following allowed values of SM\cite{PDG} quark masses
at the scale of $m_t$, where the QCD and QED RG running\cite{RG} have been
taken care of:
\beqa
m_u&=& (1.766\pm .951)\times 10^{-3} GeV\nonr\\
m_d&=& (3.26 \pm 1.63)\times 10^{-3} GeV\nonr\\
m_s&=& (62.5 \pm 29.89)\times 10^{-3} GeV\nonr\\
m_u/m_d &=& (.45 \pm .25)\nonr\\
(m_u+m_d)/2&=& (2.174\pm 1.087)\times 10^{-3} GeV\nonr\\
{m_s - (m_u+m_d)/2 \over m_d -m_u }&=& (42.5\pm 8.5)\nonr\\
m_c&=& (.576 \pm .069) GeV\nonr\\
m_b&=& (2.742\pm .097) GeV\nonr\\
m_t&=& (166 \pm 5) GeV\nonr\\
|V_{us}| &=& (.2205 \pm .0035)\nonr\\
|V_{ub}| &=& (.00315 \pm .00135)\nonr\\
|V_{cb}| &=& (.039 \pm .003)\nonr
\eeqa

\section{Feynman Rules}
In the following list we summarize
the vertices used  to carry out the analysis. The derivations including
gauge fixing procedure are given in \cite{TRIUMF}.

For coupling to the gauge bosons,
\begin{center}
\epsfig{file=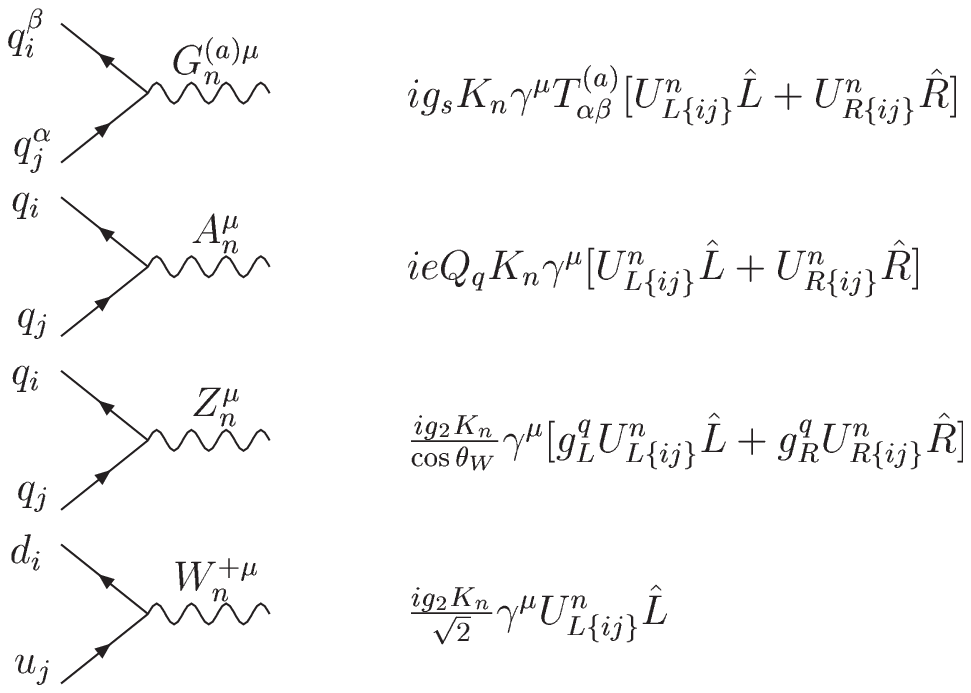, width=330pt}
\end{center}
where $i,j=1..3$ are the indices for family and $\alpha,\beta$
stand for color
with the chiral projection operators defined as
$\hat{R}/\hat{L}=\frac12(1\pm\gamma^5)$ and the standard
left-handed/right-handed $Z$ coupling,
$g_{L/R}^q= T_3(q)-Q_q \sin^2\theta_W$.
The factor $K_n$ lumps together the $\sqrt{2}$ normalization factor
for the $n$-th KK excitation and the Gaussian suppression factor,
\beq
K_n=\left\{\begin{array}{cc}
  \sqrt{2} \exp[-n^2\rho^2/4]& n>0 \\
  1& n=0
\end{array}\right.
\eeq
The mixing matrices $U^n$ are given:
\beqa
U^n_{L\{ij\}}= \sum_k (V_L^\dag)_{\{ik\}} \cos\frac{ny^L_k}{R}(V_L)_{\{kj\}},\\
U^n_{R\{ij\}}= \sum_k (V_R^\dag)_{\{ik\}}
\cos\frac{ny^R_k}{R}(V_R)_{\{kj\}}.
\eeqa
It is understood that the left/right rotation matrices $V_{L/R}$
are the ones associated with the external fermions.
For $n=0$, the cosine weighting factor reduces to one and the
diagonal cosine matrix sandwiched in between also reduces to the three
by three identity matrix. Thus, we have
$U^0_{L/R\{ij\}}=\delta_{ij}$, which are  the SM cases without FCNC
and CPV at tree level.

For the Higgs couplings, although complicated, the vertices can also
be derived straightforwardly,
\begin{center}
\epsfig{file=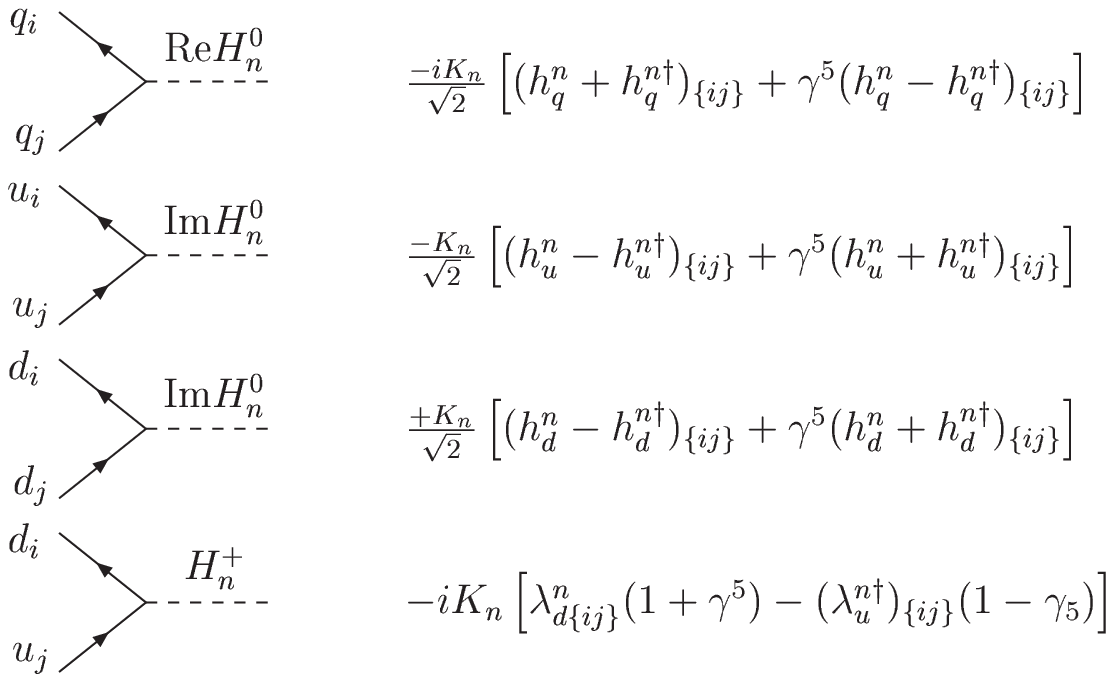, width=330pt}
\end{center}
where
\beqa
h^n_{q\{ij\}} &=& {g_2 \over \sqrt{2} M_W}\sum_{k,l=1}^3\left[(V^q_L)^\dag_{\{ik\}} M^q_{\{kl\}}
\cos \frac{n \rho(y^L_k+y^R_l)}{2}
 (V^q_R)_{\{lj\}}\right],\\
\lambda^n_{d\{ij\}}&=&{g_2 \over \sqrt{2} M_W} \sum_{k,l=1}^3\left[(V^u_L)^\dag_{\{ik\}} M^d_{\{kl\}}
\cos\frac{n\rho (y^{uL}_k+y^{dR}_l)}{2}
 (V^d_R)_{\{lj\}}\right],\\
\lambda^n_{u\{ij\}}&=& {g_2 \over \sqrt{2} M_W}\sum_{k,l=1}^3\left[(V^d_L)^\dag_{\{ik\}} M^u_{\{kl\}}
\cos\frac{n\rho (y^{dL}_k+y^{uR}_l)}{2}
 (V^u_R)_{\{lj\}}\right].
\eeqa
Again, it can be seen that when $\rho\ra0$ the cosine weighting
becomes unity  and the vertex couplings  reduce to SM cases as
expected, namely,
 $h^n_{q\{ij\}}\ra {g_2 m^q \over \sqrt{2} M_W}\delta_{\{ij\}}$,
$\lambda^n_{d\{ij\}}\ra {g_2 m^d_j \over \sqrt{2}
M_W} (V_{CKM}^\dag)_{\{ij\}}$,
 and $\lambda^n_{u\{ij\}}\ra {g_2 m^u_i \over \sqrt{2} M_W}
 (V_{CKM}^\dag)_{\{ij\}}$.

\newpage
\bibliographystyle{unsrt}

\end{document}